\documentclass[aps,prb,twocolumn,superscriptaddress,longbibliographys]{revtex4-2}

\usepackage{graphicx} 
\usepackage[utf8]{inputenc}
\usepackage{bm}
\usepackage{natbib}
\usepackage{color}
\usepackage{epstopdf}
\usepackage{amsmath}
\usepackage{amssymb}
\usepackage{bm}
\usepackage{enumerate}
\usepackage{chngpage}
\usepackage{url}
\usepackage{braket}
\definecolor{darkGreen}{RGB}{0,110,0}
\definecolor{darkBlue}{RGB}{0,0,130}
\usepackage[colorlinks=true, urlcolor=darkBlue,citecolor=darkGreen,linkcolor=darkBlue]{hyperref}

\newcommand{\bk}{{\bm{k}}}

\begin{document}


\title{Upper critical field in few-layer Ising superconductors}

\author{Lena Engstr\"om}
\email{lena.engstrom@universite-paris-saclay.fr}
\affiliation {Université Paris-Saclay, CNRS, Laboratoire de Physique des Solides, 91405 Orsay, France}
\author{Andrej Mesaros}
\affiliation{Université Paris-Saclay, CNRS, Laboratoire de Physique des Solides, 91405 Orsay, France}
\author{Pascal Simon}
\affiliation{Université Paris-Saclay, CNRS, Laboratoire de Physique des Solides, 91405 Orsay, France}

\begin{abstract}
The $N$-layer 2H-stacked transition metal dichalcogenides 2H-NbSe$_2$ and 2H-TaS$_2$ are superconductors in which each quasi-two-dimensional layer breaks inversion symmetry. In this paper, we show that, as for the individual monolayers, it is crucial to include all pockets at the Fermi surface to accurately determine the upper critical field. Furthermore, we propose an experiment where a distinct scaling with a varying displacement field is predicted for an intralayer spin-singlet order in a bilayer. The scaling of the upper critical field with external tuning parameters can thus be used to extract information about the spin-symmetry of the superconducting order. We also explore the possibility  of a mixed-parity spin-singlet and -triplet order parameter. In that case, we predict that the experimentally observable scaling would remain that of the spin-singlet component.
\end{abstract}

\maketitle

\section{Introduction}
In superconducting quasi-two dimensional (2D) layers the upper critical field $H_{c2}(T)$, the field at which the system is no longer superconducting, can reach surprisingly high values. A magnetic field couples to both the orbital motion and the spins of a superconducting Cooper pair. Coupling to the spin, via a Zeeman term, a conventional spin-singlet superconductor can reach a maximal limit $H_p = 1.76 k_{\text{B}} T_c / \sqrt{2} \mu_B$, referred to as the Pauli limit\cite{Chandrasekhar1962, Clogston1962}. However, it was long recognized that this Pauli limit can be exceeded in presence of the spin-orbit coupling (SOC), given that the field $\bm{h}$ and the SOC $\bm{g}_{\bm{k}}$ are perpendicular\cite{Klemm1975,Frigeri2004, Frigeri2004a}. The Ising superconductors are a family of superconducting transition metal dichalcogenides (TMDs) where a so-called Ising SOC (out-of-plane $\bm{g}_{\bm{k}} \propto \hat{z}$) arises in individual layers that break inversion symmetry.

In particular, the 2H-stacked TMD families 2H-NbSe$_2$ and 2H-TaS$_2$ remain superconducting from monolayer to the bulk-limit, with $H_{c2}>H_p$ for in-plane fields\cite{Xi2016, Yang2018, DeLaBarrera2018}. Focusing on the monolayer, there are clearly distinguishable signatures from different spin-triplet orders, depending on the spin-direction $\bm{d}$, while the critical field of a spin-singlet order depends directly on the SOC\cite{Frigeri2004}. Recently, a unique square-root scaling of $H_{c2}/H_p$ with the strength of the SOC for a $\Gamma$-pocket was predicted\cite{Engstrom2025}. The underlying mechanism is that pockets at the Fermi surface (FS) are spin-split, except where they intersect with symmetry-enforced nodes of the SOC. The pocket centered on the $\Gamma$-point always has such points and is hence the most important for determining the critical field\cite{Engstrom2025}.

In this work, we want to examine if a similar mechanism is present in the multi-layer 2H-stacked compounds. It is therefore necessary that we consider the full multi-layer multi-pocket systems. As the bands of 2H-NbSe$_2$ and 2H-TaS$_2$ can be described via the same model\cite{DeLaBarrera2018}, comparing the critical fields as the number of layers is increased can provide further insight into possible differences between the two compounds. For multi-layers, we must add the interlayer hopping parameters, which play an important role\cite{Maruyama2012}. In the presence of Ising SOC it is also expected that the superconducting order parameter is of mixed parity, i.e., mixing spin-singlet and -triplet symmetries\cite{Hsu2017}.

By comparing different symmetries of the superconducting order parameter in $N$ layers, we can determine that an intralayer spin-singlet order in a 3-pocket model for the upper critical field provides a qualitatively consistent description of the available experimental data. 
However, due to the multiple approximated model parameters, conclusions about the nature of superconductivity from the magnitude of the critical field become muddled. In contrast, we find that the unique scaling of the upper critical field with the SOC strength 
from an intralayer spin-singlet component found in the monolayer case extends to the few-layer systems. We show that this scaling could be extracted via experiments in a bilayer by tuning the experimentally accessible bias potential between the layers. 


This paper is structured by first introducing the model and method for calculating the upper critical field via the susceptibility in multi-pocket systems in section~\ref{sec:Model}. In section~\ref{sec:2Hbi}, we present our main result for the bilayer. We consider a bilayer both with preserved inversion symmetry and when a small symmetry-breaking term is induced by a substrate. We find that a one-component spin-singlet order can explain the experimental data for a 3-pocket model. However, the interlayer bias potential required is significant, indicating that possible subleading terms in the susceptibility should be accounted for. In section~\ref{sec:dmu} we propose an experiment for extracting a scaling with an external displacement field as a distinct signature of spin-singlet order in the bilayer. Moreover, we evaluate the effects of having a mixed-parity superconductivity, in section~\ref{sec:mixParity}, showing that the response from the spin-singlet term is the leading order. As a final point, in section~\ref{sec:N2multi} we extend the discussion for up to $N=5$ layers and show that the same qualitative behavior exists in all layers.

\begin{figure*}
    \centering
    \includegraphics[width=0.69\linewidth]{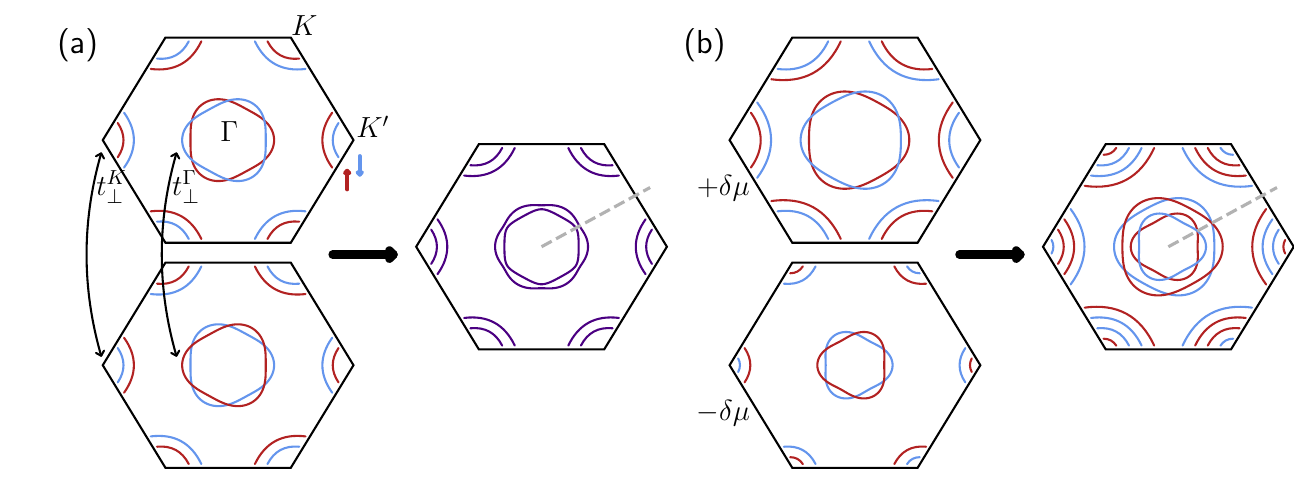}
    \includegraphics[width=0.3\linewidth]{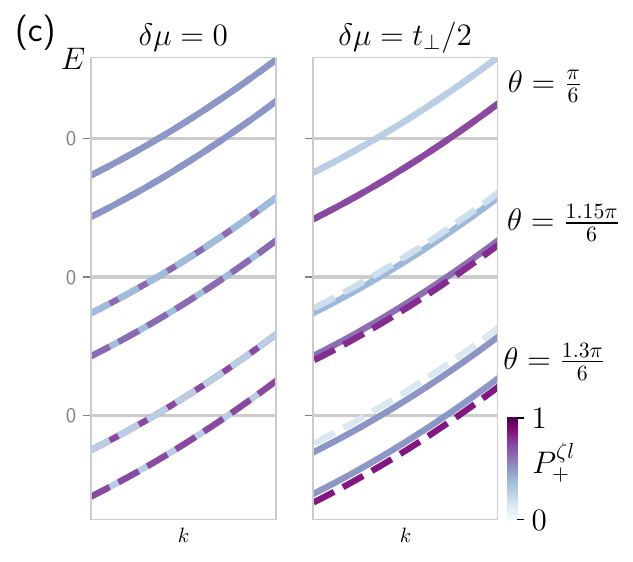}
    \caption{\label{fig:FSlayers}a) Top and bottom layers of bilayer 2H-NbSe$_2$ have 3 pockets centered around the $\Gamma$-, $K$-, and $K'$-points, which are coupled via the interlayer hopping $t_\perp^j$, $j=\Gamma,K$. The Ising SOC changes sign between the layers $\bm{g}_l \propto l \hat{z}$ , with top and bottom layer $l = \pm 1$ respectively, and its nodal points are gaped out for the degenerate bands in the bilayer. The spin-polarized spin-up (red) and -down (blue) bands are fully degenerate (purple) in the bilayer. b) When inversion symmetry is broken via an interlayer bias potential $\delta \mu$, Eq.~\eqref{eq:dmuDef}, the bands split and the SOC nodes reappear. c) Normal state band structure of a $\Gamma$-pocket around the FS (at $E=0$) close by a nodal line $\lambda_{\bm{k}}^\Gamma (\pi/6) =0$ (dashed lines in (a) \& (b)). When inversion symmetry is present, $\delta \mu=0$, the bands remain degenerate for spin-up (dashed lines) and spin-down (solid lines) even when $\theta \neq \pi/6$. The layer polarization $P_{+}^{\zeta l}$, see appendix~\ref{sec:AppEigs}, is also mixed at $\theta = \pi/6$. In contrast, for $\delta \mu>0$ the bands are only degenerate at the nodal point and have a clear layer-character close to the node. }
\end{figure*}

\section{Model}\label{sec:Model}

\subsection{Hamiltonian}\label{sec:Ham}
In the superconducting state, an external magnetic field suppresses the Cooper pairs by coupling both to the spin, via a Zeeman term, as well as the orbital motion of electrons. Of the two, only the Zeeman term is sensitive to the spin-direction of the pairs. Therefore, the relative size of the terms is of outmost importance. In few-layer 2H-NbSe$_2$ under an in-plane magnetic field, the orbital contribution has experimentally been determined to be small\cite{Matsuoka2020}. However, for bulk-like samples the orbital suppression becomes dominant\cite{Wan2023}. 

For an in-plane magnetic field along the $x$-direction the vector potential in each layer $l$ is $\boldsymbol{A}_l =  h z_l \hat{y}$, where $z_l$ is the layer's position in the $z$-direction and $h$ is the magnetic field strength. The magnetic field couples to the normal state dispersion of a band $\zeta$ as $  \xi_{\zeta l \bk} \rightarrow \xi_{\zeta l (\bk + \delta \boldsymbol{k}_l)}$. The shift in momentum is $\delta \boldsymbol{k}_l = \frac{e \boldsymbol{A}_l}{\hbar}$, with the elementary charge $e$ and Planck's constant $\hbar$. For two layers separated by a distance $d$ the difference between them is thus
\begin{equation}
    \delta \bk_{\text{top}} - \delta \bk_{\text{bottom}} = \frac{e d}{\hbar} h \hat{y}  .
\end{equation}
In the Hamiltonian for a single monolayer the Pauli matrices $\sigma_i$, $\tau_i$ are used for spin and particle-hole degrees of freedom. For multi-layers, the layer degree of freedom is introduced via the matrices $\gamma_i$, whose dimensions depend on the number of layers. For a bilayer these are Pauli matrices $i=0,x,y,z$ that act on the top ($l=+1$) and bottom ($l=-1$) layers. In 2H-stacked TMDs, the Ising SOC switches signs between the two layers and is given by $\hat{\lambda}_\bk = \bm{g}_\bk \cdot \bm{\sigma} \otimes \gamma_z  $. Hence in the 2H-stacking the overall inversion is preserved. The $\bm{g}_\bk \parallel \hat{z}$ is the SOC coupling function arising from the broken inversion symmetry within an isolated layer, respecting the crystalline symmetry. Including the Zeeman term and keeping only linear terms in the magnetic field $h$, the non-interacting Hamiltonian is:
\begin{equation}\label{eq:fullHam}
    \hat{\mathcal{H}}_n = \hat{\mathcal{H}}_{0}+ \hat{\mathcal{H}}_{\text{SOC}} +  \hat{\mathcal{H}}_{\text{inter}} + \hat{\mathcal{H}}_{\text{Z}} + \hat{\mathcal{H}}_{\text{orb}},
\end{equation}
with $\hat{\mathcal{H}}_{0}= \xi_{\bm{k}} \sigma_0 \otimes \gamma_0 $, $\hat{\mathcal{H}}_{\text{SOC}}=\lambda_{\bm{k}} \sigma_z \otimes \gamma_z $, $\hat{\mathcal{H}}_{\text{orb}}=- 2 v_{F,\bk_H} \frac{e d h}{2 \hbar} \sigma_0 \otimes \gamma_z $, $\hat{\mathcal{H}}_{\text{Z}}=\mu_{\text{B}} \bm{h}\cdot \bm{\sigma} \otimes \gamma_0$, $\hat{\mathcal{H}}_{\text{inter}}=t_\perp \sigma_0 \otimes \gamma_x $. The average Fermi velocity parallel to the magnetic field is defined as $v_{F, \bk_H}= \frac{1}{N N_\zeta} \sum_{\zeta,l} \frac{\partial \xi_{\zeta l}}{\partial k_x} $ for $N$ layers and $N_\zeta$ bands. By defining $ \mu_{\text{orb}} = \frac{v_{F, \bk_H} e d}{2 \hbar }$ the orbital contribution becomes comparable to the Bohr magneton $\mu_{\text{B}}$\cite{Zhang2023}.

When including a superconducting order we express the Hamiltonian in the Nambu space $\{ c_{\bk, \uparrow l}, c_{\bk, \downarrow l} , c_{-\bk, \downarrow l}^\dagger , - c_{-\bk, \uparrow l}^\dagger \}$, per layer $l$, with:
\begin{align}\label{eq:HamSC}
    \hat{\mathcal{H}}_s = \left( \hat{\mathcal{H}}_{0} +\hat{\mathcal{H}}_{\text{SOC}} +  \hat{\mathcal{H}}_{\text{inter}} \right) \otimes\tau_z \\ \notag + \hat{\mathcal{H}}_{\text{Z}} \otimes\tau_0 + \hat{\mathcal{H}}_{\text{orb}} \otimes\tau_z  + \hat{\mathcal{H}}_\Delta.
\end{align}
$\hat{\mathcal{H}}_\Delta$ contains the mean-field decoupled interactions into the superconducting order parameters. When including the layer degree of freedom, additional symmetries of the superconducting order parameter are allowed. The possible symmetries of the order parameter must result in an overall odd function under exchange of all labels:
\begin{equation}\label{eq:oddDe}
    \Delta_{\sigma l, \sigma' l'} (\bk) = - \Delta_{\sigma' l', \sigma l} (- \bk),
\end{equation}
for $ \Delta_{\sigma l, \sigma' l'} (\bk) = \langle c_{\bk, \sigma l}  c_{-\bk, \sigma' l' } \rangle$, as the superconducting wavefunction is formed by fermions. We consider only a center-of-mass $\boldsymbol{q}=0$ of the pairing in this work, excluding a discussion of possible Fulde-Ferrell-Larkin-Ovchinnikov  orders\cite{Wan2023}. Inherent to evaluating the susceptibility, this work makes predictions only sufficiently close to $T_c$, which makes such orders unlikely.

This paper focuses on the family of 2H-stacked TMDs which have three pockets on the Fermi surface, specifically NbSe$_2$ and TaS$_2$. The pockets are centered around the $\Gamma$-, $K$- and $K'$-points, see Fig~\ref{fig:FSlayers}. We will consider pockets of $\Gamma$- and $K$-type separately, where local parameters are defined by an expansion of a realistic model\cite{Rahn2012}. The density of states at the Fermi surface (FS) for the pockets are $N_\Gamma (0)$ and $N_K(0)$, where pockets are assumed to be of equal size $N_K(0)= 2 N_\Gamma (0)$\cite{Nakata2018}. The SOC $\lambda_{\bk}$ is defined on each pocket as $\lambda_{\bk}^K=\lambda_{0}^K$ and $\lambda_{\bk}^\Gamma=\lambda_{0}^\Gamma \cos 3 \theta$\cite{He2018}, with sizes $\lambda_0^\Gamma$ and $\lambda_0^K$ given by the band splitting found in ARPES\cite{He2018, Nakata2018}. For each pocket $j=\Gamma,K$, the dispersion is assumed to be parabolic $\xi_{\bk}$ in the range $\pm \epsilon$, with $\epsilon > \lambda_0^j$. Layers are coupled by an interlayer hopping $t_\perp$, see Fig.~\ref{fig:FSlayers}. This coupling is stronger for the $\Gamma$-pocket\cite{McHugh2023} due to the increased presence of $d_{z^2}$-orbitals\cite{Noat2015,Wickramaratne2023}. For simplicity, the interlayer hopping between each pocket $t_\perp^\Gamma$ and $t_\perp^K$ will in this work be assumed to be equal $t_\perp^\Gamma =t_\perp^K =t_\perp$.

The superconducting order parameters pair either intra-pocket electrons for $\Gamma$ or electrons between $K$- and $K'$-pockets. Transport experiments suggest that $\Delta^\Gamma$ \& $\Delta^K$ are  of equal size and  follow the same temperature dependence\cite{Kuzmanovi2022}. We will assume it to be the case and therefore  implicitly making assumptions about the pairing mechanism and coupling terms between pockets. The temperature dependence of the order parameters is modeled as the one-band mean field solution for $h=0$ with $\Delta(T=0) = \Delta_0 = 1.76 k_{\text{B}} T_c$.

\subsection{Upper critical field from susceptibility}\label{sec:Hc2Calc}
Close to the critical temperature $T_c$, the upper critical field $H_{c2}(T)$ can be considered small enough to be defined via an expansion using the spin-susceptibility. Close to $T=0$ qualitative differences in the upper critical field have been found\cite{Harms2026}, which are not present in the temperature regime considered in this work ($T \approx T_c$). For the normal state $(m=N)$ and for a proposed superconducting state $(m=s)$, the critical field is defined as the field at which the thermodynamic potentials $\Omega_m(T,H)$ are of equal value\cite{Clogston1962,Chandrasekhar1962}: 
$ \Omega_N (T, H_{c2}) -  \Omega_s (T, H_{c2}) \approx  \Omega_{0} (T) + H_{c2}^2 (\chi_N (T)- \chi_s(T) ) =0$, where we introduced the susceptibilities $\chi_m (T)$ and condensation energy $ \Omega_{0} (T)$. When considering a multi-pocket system, $\Omega^{\text{tot}}_0(T)=\sum_{j\in\{\rm pockets\}}\Omega_{0}^j(T)$ and the difference in susceptibility is $\delta \chi^{\text{tot}}(T) = \chi_N^{\text{tot}} (T)- \chi_s^{\text{tot}} (T) $. The definition of the upper critical field therefore is
\begin{equation}\label{eq:HcDef}
    H_{c2} (T) = \sqrt{\frac{\Omega^{\text{tot}}_0(T) }{\delta \chi^{\text{tot}}(T) }}.
\end{equation}
For the 2H-TMD systems we will use $ \Omega^{\text{tot}}_0(T) =  \Omega^{\Gamma}_0(T) + \Omega^{K}_0(T)$ and $\delta \chi^{\text{tot}}(T) = \delta \chi^{\Gamma}(T) + \delta \chi^{K}(T)$. The susceptibility for each pocket can be divided into intra- and inter-band terms\cite{Sigrist2009,Samokhin2021}
\begin{align}
    \chi_m (T) = \left. \frac{\partial^2 \Omega_m (T,h)}{\partial h^2} \right|_{h=0} = \chi_m^{\text{intra}}(T)  + \chi_m^{\text{inter}} (T)
\end{align}
with $m=N,s$. 

The Pauli limit is defined as the critical field found via Eq.~\eqref{eq:HcDef} with a purely Zeeman coupling for singlet superconductivity in a single band, $\chi_N (T)=\chi_P$, and $\chi_s (T=0)=0$. In general, the size of $H_{c2}$ is determined by the leading term $\delta \chi_n(T) = \mathcal{O}(\Delta^n)$. Furthermore, the scaling of $\delta \chi(T) $ with the model parameters can distinguish different types of pairing symmetries. In particular, for a monolayer with a $\Gamma$-pocket only a singlet pairing results in a $H_{c2} \propto \sqrt{\lambda_0^\Gamma}$ scaling for an in-plane field ($\bm{h} \perp \bm{g}_{\bm{k}}$)\cite{Engstrom2025}.

\section{Upper critical field in 2H bilayers}\label{sec:2Hbi}

\subsection{Susceptibility for singlet pairing}

In this section we consider the pairing to be a 2D spin-singlet within each layer: $\hat{\Delta}= \Delta \sigma_0  \otimes \gamma_0 \otimes \tau_x$. The magnitude is assumed to be equal on all pockets. However, we will see how this assumption is less relevant here than in the monolayers, where the ratio $\Delta_0^j /\lambda_0^j$ on each pocket $j= \Gamma,K$ is of importance\cite{Engstrom2025}. 

Using model parameters for few-layer 2H-stacked TMD we can compare the size of the terms in Eq.~\eqref{eq:fullHam}, to determine whether the orbital coupling is negligible compared to the Zeeman coupling. For a bilayer NbSe$_2$ $v_F \approx 0.55 \times 10^{6}$ m/s at the $\Gamma$-pocket and $d = 0.7 \times 10^{-9}$m\cite{DeLaBarrera2018}. We thus approximate $\mu_{\text{orb}}=v_F \frac{e d}{2 \hbar} \approx 0.007 \mu_{\text{B}} $ and $\delta \chi^{\text{orb}} \propto \mu_{\text{orb}}^2\approx 5\times 10^{-5}\mu_{\text{B}}^2$. This can be compared to the monolayer Zeeman term\cite{Engstrom2025}, where if $\frac{\lambda_0}{\Delta_0} \approx 17$ then $\delta \chi^{\text{Z}} \propto  \frac{\Delta_0}{\lambda_0} \frac{\pi}{\sqrt{3}} \mu_{\text{B}}^2 \approx 0.1 \mu_{\text{B}}^2 \gg \delta \chi^{\text{orb}} $. For each layer the normal state susceptibility can thus easily be calculated using only a Zeeman coupling of the field to the spins. 

For intralayer spin-singlet superconductivity, the susceptibility can never reach  values beyond the normal state interband susceptibility\cite{Maruyama2012, Sigrist2014, Skurativska2021}, as the remaining intraband susceptibility originates from the Fermi surface. In a bilayer 2H-TMD the SOC nodes disappear and instead two sets of degenerate bands are separated by a distance proportional to $t_\perp$, see Fig.~\ref{fig:FSlayers}. (Recall, in our model we take the same $t_\perp$ for all pockets.) The now leading term  $\delta \chi^j_0 = \mathcal{O}(\Delta^0)$ is determined by the ratio $t_\perp/ \lambda_0^j$. For $K$- and $\Gamma$-pockets respectively, we find (see appendix~\ref{sec:AppSuscp}):
\begin{align}\label{eq:2LKG}
   \delta \chi^{K,\text{2L}}(T) \approx  \frac{ t_\perp^2}{ t_\perp^2 + \left(\lambda_0^K\right) ^2} (\chi_P - \chi_{\text{sg}}(T) ),\\
   \delta \chi^{\Gamma,\text{2L}}(T) \approx \frac{ t_\perp }{ \sqrt{ t_\perp^2 + \left(\lambda_0^\Gamma \right) ^2} } (\chi_P - \chi_{\text{sg}}(T) ),
\end{align}
where $ \chi_{\text{sg}}(T) = Y(T) \chi_P$ and $ Y(T)$ is the Yoshida function\cite{Sigrist2009} (see appendix~\ref{sec:AppSuscp}). For any number of layers $N$ the leading term is $\delta \chi^j_0 = \mathcal{O}(\Delta^0)$, as detailed in section~\ref{sec:N2multi} for up to $N=5$ layers.

\begin{figure}
\includegraphics[width=\linewidth]{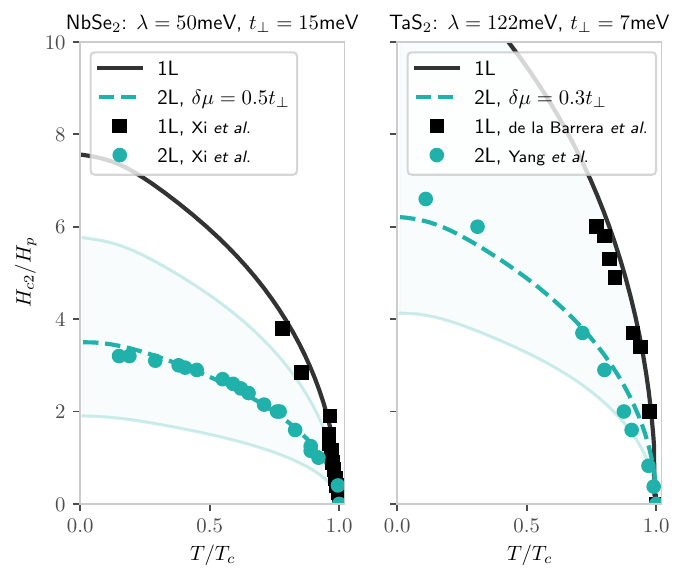}
\caption{\label{fig:biLayer}In-plane upper critical field for mono- and bi-layer NbSe$_2$\cite{Xi2016} and TaS$_2$\cite{DeLaBarrera2018,Yang2018}. For bilayers a fit for the interlayer bias potential $\delta \mu$ is shown. For multi-layers a lower bound of $H_{c2}(T)$ is Eq.~\eqref{eq:2LKG}, when it is suppressed by hopping between the layers at $\delta \mu =0$. The upper bound is achieved when the layers are treated as uncoupled monolayers at the given $T_c$ ($\delta \mu \rightarrow \infty$).}
\end{figure}

To deal with variation due to multiple model parameters, we provide an upper and lower bound for the $H_{c2}$ in Fig.~\ref{fig:biLayer} with $\delta \chi^{\text{tot}}(T) = \delta \chi^{\Gamma,\text{2L}}(T) + 2 \delta \chi^{K,\text{2L}}(T) \approx \delta \chi^{\Gamma,\text{2L}}(T) $. The lower bound is given by Eq.~\eqref{eq:2LKG}, where the sample has no interlayer bias potential, e.g., a free-standing sample. The upper bound is the limit where each layer is treated as an isolated monolayer.

To illustrate the necessity of including all three pockets, the lower bounds for the 3-pocket model, in Fig.~\ref{fig:biLayer}, should be compared to those found when only considering the $K$-pockets, see appendix~\ref{sec:dissVal}. The 3-pocket model qualitatively describes the experimental data for both 2H-NbSe$_2$ and 2H-TaS$_2$, as the lower bounds underestimate the critical field by roughly the same amount. In contrast, the $K$-pocket model greatly overestimates the critical field for bilayer TaS$_2$.

\subsection{Effective monolayer model}\label{sec:effMono}
Another non-negligible factor in the multilayer case is a small inversion symmetry breaking coming from a substrate. In the bilayer, we hence include an interlayer bias potential\cite{Samuely2023}, $\delta \mu$, as
\begin{equation}\label{eq:dmuDef}
    \hat{\mathcal{H}}_{\delta\mu} = \delta \mu  \sigma_0 \otimes \gamma_z \otimes  \tau_z,
\end{equation}
in the Hamiltonian in Eq.~\eqref{eq:fullHam}, setting $\mu_{\text{orb}}=0$.

The normal state bands are fully spin-polarized and mix the layer degrees of freedom, see appendix~\ref{sec:AppEigs}. The interlayer bias potential $\delta \mu$ shifts the chemical potential of the two layers in opposite direction. As the states of the two layers no longer overlap at the FS (see Fig.~\ref{fig:FSlayers}b), the mixing of the layer degrees of freedom, due to the interlayer hopping, is decreased. The bands, which were two-fold degenerate at $\delta \mu=0$, see Fig.~\ref{fig:FSlayers}c, are now spin-split:
 \begin{equation}\label{eq:eigDmu}
     \xi_{\zeta l} = \xi_{\bk} + \zeta l \sqrt{ \left(\zeta l|\lambda_{\bk}| + l \delta \mu \right)^2 + t_\perp^2 },
 \end{equation}
 at all points expect at the SOC nodes $\lambda_{\bm{k}}=0$ when $\delta \mu \neq 0$. The label $\zeta = \pm 1$ is the spin and $l= \pm 1$ is the layer in which the state is primarily located. As can be seen in Fig.~\ref{fig:FSlayers}c, close to the SOC nodes the bands have a clear layer-character.

Following the conclusions from the monolayer case\cite{Engstrom2025}, the susceptibility for an intralayer singlet superconducting order is almost entirely determined by the states close to the SOC nodes, where the effective SOC at any momentum point should be the splitting between states of opposite spins within the same layer. The inversion-symmetry-broken bilayer can therefore be modeled as two monolayer bands with an effective SOC $\tilde{\lambda}_0$, as we describe now.
 
Without an applied magnetic field $h$ the bands in the superconducting state are $E_{\zeta l} = \sqrt{\xi_{\zeta l}^2 + |\Delta|^2}$. The definition of the effective SOC should therefore come from the normal state band splitting close to a SOC node. The expansion in small $|\lambda_{\bk}|$ of Eq.~\eqref{eq:eigDmu} results in the bands
  \begin{equation}\label{eq:eigsDmuLa}
     \xi_{\zeta l} \approx \xi_{\bk} + l \sqrt{ \delta\mu^2 + t_\perp^2 } + \zeta  l \frac{\delta \mu |\lambda_{\bm{k}}|}{\sqrt{ \delta \mu^2 +t_\perp^2}}.
 \end{equation}
For two bands with opposite spin $\zeta$ and the same layer polarization $l$, the splitting is thus $2 \frac{\delta \mu |\lambda_{\bm{k}}|}{\sqrt{ \delta \mu^2 +t_\perp^2}}$. The effective SOC is defined as
\begin{equation}\label{eq:effLaBi}
    \tilde{\lambda}^j_0 = \frac{\delta \mu}{\sqrt{\delta \mu^2 + t_\perp^2}} \lambda^j_0.
\end{equation}
If $\delta \mu$ is sufficiently large, the critical field thus scales as an effective monolayer model \cite{Engstrom2025}
\begin{equation}
H_{c2}^{2L} /H_p^{2L} \propto \sqrt{\frac{ \tilde{\lambda}^\Gamma_0}{\Delta_0^{\Gamma,\text{2L}}}}.
\end{equation}
More generally, compared to a single monolayer, the bilayer $H_{c2}$ depends on an increased number of parameters: $\lambda^j_0, \Delta_0^{j,\text{2L}}, t_\perp, \delta \mu$ for each pocket $j$. In Fig.~\ref{fig:biLayer} the value $\delta \mu$ has been fitted to experimental data, while $\lambda^j_0, t_\perp$ are given by Ref.~\cite{DeLaBarrera2018}, and $\Delta_0^{j,\text{2L}}$ by $T_c$ for the fitted dataset. For both NbSe$_2$ and TaS$_2$ we find $\delta \mu < 10$meV, which could be a reasonable, albeit somewhat high, value\cite{Dreher2021,Hall2019,Samuely2023,Zullo2023}. 



\section{Interlayer bias potential}\label{sec:dmu}
\begin{figure}
    \centering
    \includegraphics[width=\linewidth]{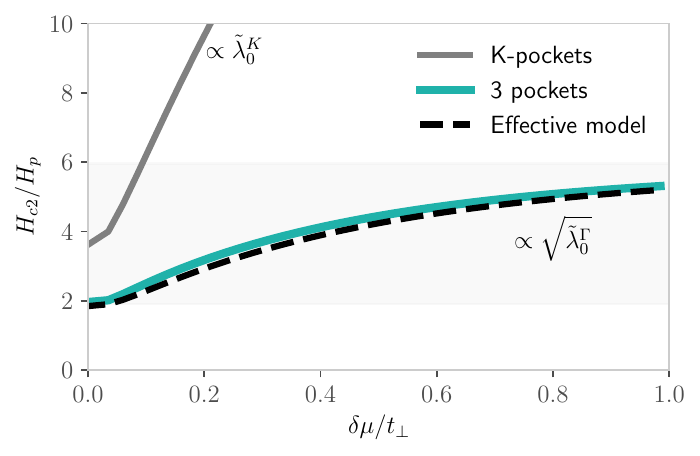}
    \caption{\label{fig:dmu_var}Prediction for bilayer 2H-NbSe$_2$ with increasing $\delta \mu$ evaluated numerically, for $\lambda_0^\Gamma=17$meV, $t_\perp =15$meV, \& $T_c = 5$K, for only Zeeman coupling of the magnetic field to an intralayer singlet pairing. The effective model, section~\ref{sec:dmu}, shows that including all 3 pockets results in a scaling roughly as $\sqrt{\delta \mu / t_\perp}$ and while for only the $K$-pockets the scaling is $\delta \mu / t_\perp$.}
\end{figure}
We have introduced the non-zero interlayer bias potential $\delta \mu$ originating from a substrate, when dealing with the 2H-stacked homobilayers. As utilized to great success in Moiré graphene and TMD systems $\delta \mu$ can be tuned by varying a displacement field perpendicular to the layers\cite{Liu2017,Ye2012}. We propose that the scaling of $H_{c2}$ with $\delta \mu$ can be used to infer properties of the superconducting order.

\subsection{Scaling with SOC}
The use of a monolayer model with an effective SOC, as in Eq.~\eqref{eq:effLaBi}, is valid when $\delta \mu$ is sufficiently large. In the opposite limit, $\delta \mu \rightarrow 0$, the approximation of treating the layers separately becomes problematic, as the critical field must reach the value of when inversion symmetry is preserved, $H_{c2} = H_{c2}^{\delta \mu=0}$. To complete the analysis, we derive the behavior at small $\delta \mu$ from the full Hamiltonian.

To consider the full Hamiltonian with an intralayer singlet order we have performed a numerical integration of the susceptibility to calculate the critical field when all parameters are of intermediate size. In Fig.~\ref{fig:dmu_var} the resulting critical field is shown as a function of $\delta \mu / t_\perp$ for a model including all three pockets and a model only including the $K$-pockets. The 3-pocket model is limited by the same upper (at $\delta \mu \rightarrow \infty$) and lower (at $\delta \mu =0$) bounds of $H_{c2}$ as shown in Fig.~\ref{fig:biLayer}.

For comparison, Fig.~\ref{fig:dmu_var} shows an interpolated effective model between the $\delta \mu= 0 $ limit with Eq.~\eqref{eq:2LKG} and the effective monolayer model for $\delta \mu \rightarrow \infty $ with Eq.~\eqref{eq:effLaBi}. Details are given in appendix~\ref{sec:AppInter}. This effective model allows us to extract the scaling of the critical field with $\delta \mu / t_\perp$ at arbitrary $\delta \mu $. In analogy to the monolayer-case, the $H_{c2}/H_p \propto \sqrt{\tilde{\lambda}^\Gamma_0} \propto \sqrt{\delta \mu / t_\perp}$-scaling can only originate from spin-singlet pairing and is therefore a signature of the spin-symmetry of the pairing. 

\subsection{Proposed experiment}\label{sec:Exp}
In an experimental set-up, applying a displacement field results in an interlayer bias of the order of meV\cite{Zhu2025}. For the predicted critical field in Fig.~\ref{fig:dmu_var}, the scaling originating from an intralayer spin-singlet pairing would thus be observable as the interlayer hopping is of size $t_\perp=15$meV. As the hopping is the main parameter influencing the feasibility of observing the scaling, a smaller value, such as in 2H-TaS$_2$, could enhance the visibility of the scaling at lower $\delta \mu$. 

One note about $H_{c2}$ when varying $\delta \mu$ is that the effects from doping the bands are present. States belonging to each layer experience an opposite chemical potential shift \cite{Xi2016a,Engstrom2025}. Fortunately, in the range $\delta \mu < \lambda_0^\Gamma$, the effects such as the change in the density of states at the FS and the change in the SOC at each pocket, are negligible. The assumption in Fig.~\ref{fig:dmu_var} is thus that $T_c$ remains constant, as the value of $H_{p}$ must be scaled with the experimental $T_c$ at each point.

\section{Mixed parity}\label{sec:mixParity}
In previous sections we have considered a purely spin-singlet pairing. The broken inversion symmetry in each layer is however significant, as it causes the Ising SOC. Ising superconductivity imposes additional conditions on the pairing spin-symmetry from the combination of inversion symmetry-breaking and the preserved time-reversal symmetry of the superconducting order parameter. In general, the spin-splitting of the pockets at the FS, around the $K$- and $K'$-points, into inner and outer  pockets\cite{Wickramaratne2020}, only allows for two order parameters $\Delta_i^{\uparrow \downarrow} (\bk )$ and $\Delta_o^{\uparrow \downarrow} (\bk )$(where the indices $i$ and $o$ stand for inner and outer respectively) that are required to be
\begin{equation}
    \Delta_o^{\uparrow \downarrow} (\bk) = \langle c_{K,o, \uparrow} c_{K', o, \downarrow} \rangle, \qquad \Delta_i^{\downarrow \uparrow} = - \langle c_{K,i, \downarrow} c_{K', i, \uparrow} \rangle
\end{equation}
Because of the spin-polarization, the order parameter in each band is neither purely spin-singlet $\Delta_s$ nor -triplet $\Delta_t$:
\begin{equation}
    \Delta_i = \frac{\Delta_s - \Delta_t}{\sqrt{2}}, \qquad \Delta_o= \frac{\Delta_s + \Delta_t}{\sqrt{2}}
\end{equation}
The spin-triplet order parameter has a spin-symmetry $\bm{d} \parallel \bm{g}_{\bm{k}}$\cite{Yoshida2014}. As long as $\Delta_o \neq \Delta_i$ we expect a non-zero spin-singlet and -triplet mixing ratio. Due to the symmetry (Eq.~\eqref{eq:oddDe}) for intralayer orders, with no additional labels to be exchanged, the spin-singlet and -triplet components must be even- and odd-parity in momentum space, respectively.

\subsection{Monolayer}
\begin{figure}
    \centering
    \includegraphics[width=\linewidth]{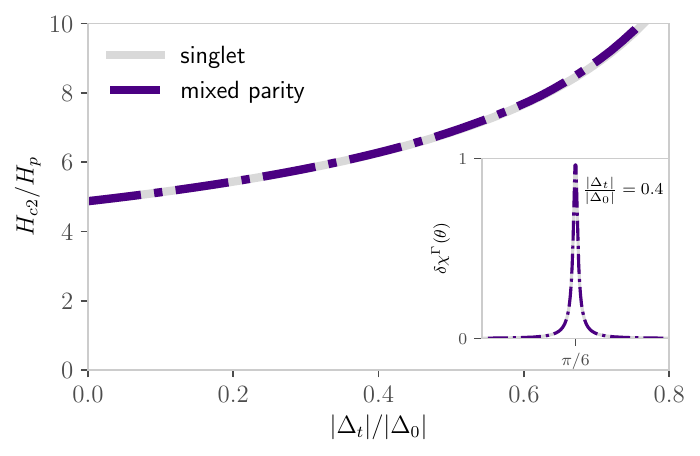}
    \caption{\label{fig:mixParity}In a one-pocket model of monolayer 1H-NbSe$_2$ with a mixed parity $s+f$-wave order, the critical field increases along with  the proportion of triplet component ($|\Delta_{s,0}| + |\Delta_{t,0}| = |\Delta_{0}|$ is kept constant). Compared to the calculation (dashed line) is that of a purely singlet order (solid line) with the same-size singlet component $H_{c2}/H_p \propto \sqrt{\lambda_0^\Gamma / \Delta_{s,0}}$, where $T_c$ is equal for both curves. The overlap of the curves shows how the increase comes entirely from the decreased singlet component and that all remaining terms, arising from the triplet component, are negligible. In the inset, the susceptibility difference at an angle $\theta$ around the $\Gamma$-pocket (SOC node at  $\theta=\pi/6$) is shown with and without the triplet component.}
\end{figure}
We begin by considering a mixed-parity order in a single monolayer. We assume that the momentum dependence of the triplet order parameter follows the Ising SOC, namely $\Delta_{t,\bk} (T) = \Delta_{t}(T) \cos 3 \theta$.
For a single $K$-pocket the triplet pairing can be considered constant $\Delta_{t,\bk}^K (T) = \Delta_{t}^K(T)$. The possible order compatible with this Fermi surface anisotropy is an $s+f$-wave type order\cite{Wickramaratne2020}. As the triplet component has coinciding nodes with the SOC only a singlet order is present at those points.
 
For real-valued order parameters the singlet-triplet mixed state has bands
\begin{equation}
    E_{\zeta}^{\text{mix}} = \sqrt{\xi_{\bk}^2 + |\lambda_{\bm{k}}|^2 + |\Delta_s|^2 + |\Delta_t|^2 + 2 \zeta  | \Delta_s \Delta_t + \xi_{\bk}\lambda_{\bm{k}}| },
\end{equation}
 Adding the Zeeman term (with $\mu_{\text{B}}=1$ for now), the bands become
\begin{equation}
    E_{h, \zeta}^{\text{mix}} = \sqrt{\xi_{\bk}^2 + |\lambda_{\bm{k}}|^2 + |\Delta_s|^2 + |\Delta_t|^2 + h^2 + 2 \zeta \gamma_{h, \bk}},
\end{equation}
with $\gamma_{h, \bk} =  \sqrt{ h^2 \left( \xi_{\bk}^2 + |\Delta_s|^2 \right) + \left( \Delta_s \Delta_t + \xi_{\bk}\lambda_{\bm{k}} \right)^2 } $ for $\bm{h} \parallel \hat{x}$. The susceptibility originates entirely from interband contributions, see section~\ref{sec:Hc2Calc}, and is calculated via: 
\begin{equation}\label{eq:mixMonoDer}
    \left. \frac{\partial^2 E_{h, \zeta}^{\text{mix}}}{\partial h^2} \right|_{h=0} = \zeta \mu_{\text{B}}^2 \frac{\xi_{\bk}^2 + |\Delta_s|^2 + \zeta | \Delta_s \Delta_t + \xi_{\bk} \lambda_{\bm{k}} | }{ | \Delta_s \Delta_t + \xi_{\bk}\lambda_{\bm{k}} | E_{\zeta}}.
\end{equation}
Where $E_{\zeta}= \left. E_{\zeta}^{\text{mix}} \right|_{\Delta_t =0}$ are the bands at $h=0$ for purely spin-singlet pairing. We consider the case that $T_c$ is known for the system. However, instead of considering one order parameter $\Delta_0 = 1.76 T_c k_{\text{B}}$ the mixed-parity order is of size $|\Delta_{s,0}| + |\Delta_{t,0}| = |\Delta_{0}|$.  A mixed-parity order with an increased proportion of triplet superconductivity is thus associated with a decreased singlet order. 

In Fig.~\ref{fig:mixParity} the critical field of the mixed-parity state is shown when the proportion $|\Delta_{t,0}| / |\Delta_0|$ increases, using a numerical integration of the susceptibility from Eq.~\eqref{eq:mixMonoDer}. The critical field increases significantly with a larger triplet proportion. In appendix~\ref{sec:AppMixed} an expansion in small $\Delta_{t,0}/\Delta_{s,0} \ll 1$ of Eq.~\eqref{eq:mixMonoDer} shows that any term originating directly from the triplet pairing is negligible in the susceptibility as long as $|\lambda_0^\Gamma| > |\Delta_{t,0}| $. 

The increase of the critical field instead originates from the smaller singlet component $|\Delta_{s,0}| < \Delta_0$. Overlapping with the result for the mixed parity Fig.~\ref{fig:mixParity} is a calculation for a monolayer singlet where $T_c$ remains constant and with $H_{c2}/H_p \propto \sqrt{\lambda_0^\Gamma / \Delta_{s,0}}$. For a mixed-parity superconducting state, we therefore predict that  $H_{c2}/H_p$ will scale with the SOC as that of a singlet order parameter, even if there is a large triplet component.

The implication for the pairing symmetry in a monolayer is that a scaling showing spin-singlet pairing at the $\Gamma$-pocket is only indicating the presence of a singlet order. However, the value of $H_{c2}/H_p$ is what can determine whether the pairing is predominantly singlet or not. For example, Fig.~\ref{fig:mixParity} tells us that $|\Delta_t| / |\Delta_0| =0.75$ would double the critical field and as seen in Fig.~\ref{fig:biLayer} the monolayer data is well-described by a purely singlet order. Realistically, a small but finite triplet component could be present. If the effect is of the same size as that of a Rashba SOC or disorder\cite{Mockli2020,Ilic2017, Engstrom2025}, which actually decreases the critical field, the two effects could ultimately cancel. As an example, a small Rashba term $\lambda_R^\Gamma/\lambda_I^\Gamma =0.1$ decreases $H_{c2}$ by the same amount\cite{Engstrom2025} by which a mixed-parity proportion $|\Delta_t| / |\Delta_0| =0.1$ increases it.

\subsubsection{Bilayer treated with an effective monolayer model}\label{sec:effMonoBiMix}
In the bilayer case this type of order parameter has little effect on $H_{c2}$ when inversion symmetry is preserved, as will be shown further in the next section. However, when $\delta \mu \neq 0$, the effective monolayer model would have a smaller singlet component for a mixed-parity order. The scaling would be as in Sec.~\ref{sec:dmu}, only now with $H_{c2}/H_p \propto \sqrt{\tilde{\lambda}_0^\Gamma / \Delta_{s,0}}$ where $\tilde{\lambda}_0^\Gamma$ is the effective Ising SOC on the $\Gamma$-pocket, as defined in Eq.~\eqref{eq:effLaBi}.
For a mixed parity order, a smaller $\delta \mu $ is thus required to fit the experimental data. For example, if $|\Delta_{s,0}| / |\Delta_0| = 0.8$ then a smaller $\delta \mu / t_\perp \approx 0.38$ reproduces the data for bilayer 2H-NbSe$_2$ in Fig.~\ref{fig:biLayer}. Even though this is a significant decrease, the value is still a considerate $\delta \mu \approx 5.7$meV.

\subsection{Bilayer}
\begin{figure}
    \centering
    \includegraphics[width=\linewidth]{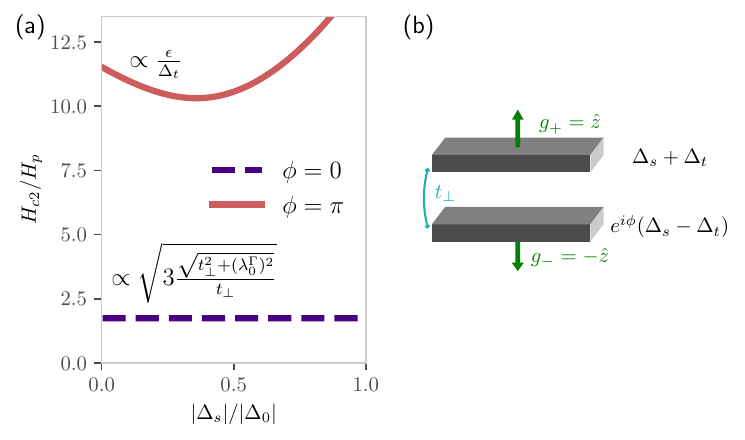}
    \caption{\label{fig:mixBi}a) For a 2H-NbSe$_2$ bilayer with inversion symmetry ($\delta \mu =0$) two intralayer mixed parity orders are shown as singlet proportion increases. When the phase $\phi=0$, see b), the critical field is dominated by the interlayer hopping and is not affected by the size of either order parameter. When $\phi=\pi$ the interlayer hopping does not suppress the critical field. }
\end{figure}
For a bilayer system there are multiple new superconducting symmetries allowed due to the added layer index $l=\pm1$ of the order parameter (see Eq.~\eqref{eq:oddDe}). In this work we consider only intralayer orders. An interlayer superconducting order parameter $\Delta_{\sigma l, \sigma' \bar{l}} (\bk)$ is possible in theory and could be either layer-singlet or -triplet. However, studies on few-layer TMDs find that neither electron-phonon coupling nor pairing mediated via spin-fluctuations favor such orders\cite{Das2023,Roy2025}.

In an inversion-symmetric bilayer, each layer taken independently breaks inversion symmetry and a spin-triplet component is thus allowed with $\bm{d}_l \parallel \bm{g}_{\bm{k},l}$:
\begin{equation}
    \hat{\Delta}_{l} \propto \hat{\Delta}_s (\bk) + l \hat{\Delta}_t (\bk) .
\end{equation}
 However, globally the superconducting order parameter should also respect inversion symmetry. As described in \cite{Maruyama2012,Liu2017}, this leaves two options for a mixed-parity order: either $\phi=0$ or $\phi=\pi$ for the order parameters: $\hat{\Delta}_{\text{top}} = \hat{\Delta}_s (\bk) +\hat{\Delta}_t (\bk)$ and $\hat{\Delta}_{\text{bottom}} = (\hat{\Delta}_s (\bk) - \hat{\Delta}_t (\bk)) e^{i \phi} $.

Explicitly, this means that for $\phi=0$ the singlet pairing has the same sign on both layers while the triplet order has opposite sign. For $\phi=\pi$ the opposite is true. Because of the phase difference $\phi$, an interlayer coupling between the layers, via the hopping $t_\perp$, will have a different effect in the two cases that we detail below.

\subsubsection{$\phi =0$}
The case of an intralayer singlet with the same phase was used in sections~\ref{sec:2Hbi} \&~\ref{sec:dmu}. As discussed in those sections, and as proposed by Ref.~\cite{Maruyama2012}, the critical field will be limited by the decreased interband susceptibility in the normal state due to the interlayer hopping. By considering the mixed state we can show that this is the leading term at any mixing ratio, and that any subleading term can be neglected.

For a spin-singlet and a layer-dependent $d_z$-triplet, the superconducting bands are
\begin{equation}\label{eq:biMixBands}
     E_{l\zeta}^{\phi=0} = \sqrt{\xi_{\bk}^2 + \lambda_{\bk}^2 + t_\perp^2 + |\Delta_s|^2 + |\Delta_t|^2 + 2 \zeta l \gamma_{\bk}^{\phi=0}},
\end{equation}
with $\gamma_{\bk}^{\phi=0}= \sqrt{(\Delta_s  \Delta_t +\lambda_{\bk} \xi_{\bk} )^2+t_\perp^2 \left(\Delta_t ^2+\xi_{\bk} ^2\right)}$ at $h=0$. A full numerical integration of the susceptibility for an increasing triplet proportion in an $s+f$-wave pairing is shown in Fig.~\ref{fig:mixBi}. Any subleading term is negligible to the point where the mixing proportion does not affect the found $H_{c2}$, even if the state is fully a triplet order. As we are considering a regime where the ratio $\frac{t_\perp }{ \lambda_0^\Gamma } \gg \frac{\Delta_s }{ \lambda_0^\Gamma }, \frac{\Delta_t }{ \lambda_0^\Gamma } $, the susceptibility in the superconducting state can be expanded as:
\begin{equation}
    \chi_s (T=0) = \int d \xi_{\bm{k}} d \theta \left( \frac{t_\perp^2}{t_\perp^2 + \lambda(\theta)^2} + \mathcal{O} (\Delta_j^2) \right).
\end{equation}

The scenario where a triplet order would have the largest impact on the critical field is when the superconducting nodes are not coinciding with the SOC nodes. For example, if the mixed-parity order has a $d+id$-wave and $p+ip$-wave symmetry\cite{Hsu2017}. In that case, the triplet component is non-zero for the states close to $\lambda(\theta=\theta_0)=0$. By expanding the susceptibility in orders of a small spin-triplet pairing $\Delta_t$ we find
\begin{equation}
    \delta \chi (\theta=\theta_0) = 1 + \frac{\Delta_t^2}{\epsilon^2} - \frac{\Delta_t^2}{t_\perp^2},
\end{equation}
as long as $\frac{\Delta_t^2}{t_\perp^2} < 1$. The full susceptibility difference is of order $\delta \chi_{\phi=0}^{d+p} \approx  \frac{t_\perp}{\sqrt{t_\perp^2 + (\lambda_0^\Gamma)^2}} \left( 1 - \frac{\Delta_t^2}{t_\perp^2} \right)$
while $\delta \chi_{\phi=0}^{s+f} \approx  \frac{t_\perp}{\sqrt{t_\perp^2 + (\lambda_0^\Gamma)^2}} $. Hence, a non-zero triplet component at the SOC node only affects the susceptibility with a correction of the size $\frac{\Delta_t^2}{t_\perp^2}$.

The state $\phi=0$ is the most likely order, as no signs of a phase shift between the layers has been established. However, as shown in this section, an increased critical field compared to the spin-singlet order in Fig.~\ref{fig:biLayer} cannot be a direct result of an intralayer spin-triplet component, like in the monolayer case.

\subsubsection{$\phi =\pi$}\label{sec:phiPi}
For the mixed state with a phase shift $\phi =\pi$ the bands $E_{\zeta l}^{\phi =\pi}$ are that of Eq.~\eqref{eq:biMixBands}, with the singlet and triplet order parameters exchanged. The major difference appears once an in-plane field is applied, as $t_\perp$ no longer suppresses the critical field. The susceptibility originates from interband terms via
\begin{align}
   \left. \frac{\partial^2 E_{h, \zeta l}^{\phi =\pi}}{\partial h^2} \right|_{h=0} & =  \zeta l  \frac{1 + \zeta l \frac{  \left(\Delta_s ^2+\xi_{\bk} ^2\right)  \left(\Delta_s \Delta_t +\xi_{\bk} \lambda_{\bk} \right)^2}{(\gamma_{\bk}^{\phi =\pi})^3} }{
   E_{\zeta l}^{\phi =\pi}} \\ & \notag +    \frac{t_\perp^2 \left(\xi_{\bk} ^2 + \Delta_s ^2 + \zeta l \gamma_{\bk}^{\phi =\pi} \right)^2}{(\gamma_{\bk}^{\phi =\pi})^2 ( E_{\zeta l}^{\phi =\pi})^3}.
\end{align}
The resulting critical field for this type of mixed state will have contrasting behavior depending on whether the dominant order is singlet or triplet. In both limits $\delta \chi \propto \Delta_j^2$ and as seen in Fig.~\ref{fig:mixBi} the resulting critical field exceeds that of a monolayer spin-singlet.

Such a result can be understood by considering a $\phi =\pi$ purely spin-singlet order $\Delta_0 = \Delta_{s,0}$ with a small interlayer hopping. The critical field for the intralayer singlet order depends strongly on the SOC within the layer that it is located in and is primarily affected by the SOC nodes at the $\Gamma$-pocket. When $t_\perp$ is non-zero it mixes the bands, and therefore the opposite sign singlets of the two layers, which cancels contributions to the susceptibility difference in each layer. An increased $t_\perp$ thus greatly increases $H_{c2}$. In the other limit, where the $\phi =\pi$ is fully spin-triplet $\Delta_0 = \Delta_{t,0}$, the critical field is mostly unaffected by interlayer hopping and instead reaches the value which is found for a spin-triplet in a monolayer.

The mixed-parity $\phi =\pi$ bilayer order might be a way to explain a higher critical field than the intralayer spin-singlet order provides. However, the values found for $H_{c2}/H_p$ pose the opposite problem, where they are far too large to explain the experimental data. Furthermore, this type of order has been predicted to only be energetically favorable for high magnetic fields in Rashba SOC compounds\cite{Chen2019, Yoshida2014}. It can thus be ruled out as a likely order, independent of the dominant spin-symmetry.

\section{2H multi-layers with $N>2$}\label{sec:N2multi}
\begin{figure}
    \includegraphics[width=\linewidth]{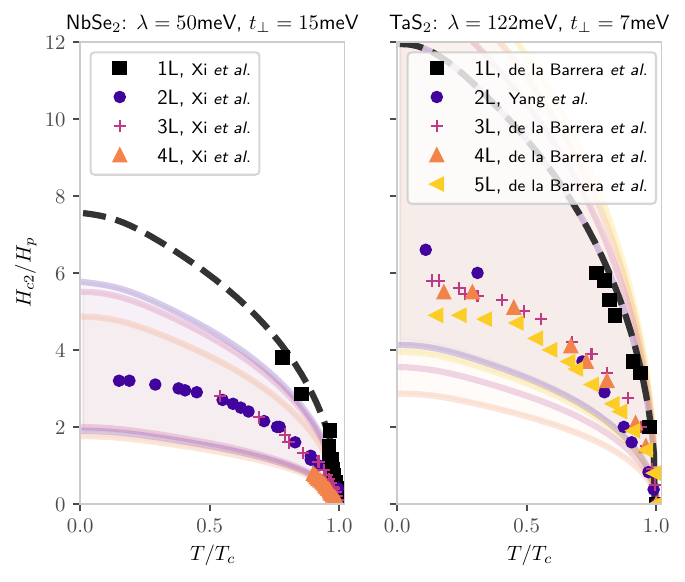}
    \caption{\label{fig:allLayers}Upper and lower bounds of $H_{c2}(T)$ for mono- and multi-layer NbSe$_2$ \cite{Xi2016} and TaS$_2$ \cite{DeLaBarrera2018,Yang2018}. The upper bounds are determined by the critical temperatures of the datasets. For NbSe$_2$: $T_c^{1L} = 3$K, $T_c^{2L} = 5$K, $T_c^{3L} = 5.3$K, \& $T_c^{4L} = 6$K. For TaS$_2$: $T_c^{1L} = 3$K, $T_c^{2L} = 3$K, $T_c^{3L} = 2.5$K, $T_c^{4L} = 2.2$K, \& $T_c^{5L} = 2$K.}
\end{figure}
2H-stacked TMDs with $N>2$ layers share the same qualitative features as the bilayers. Recall, as was calculated in section~\ref{sec:2Hbi}, the leading term of the susceptibility difference for an intralayer singlet in the bilayer is of order $\delta \chi_0$ and is given by the normal state intraband susceptibility (when $\delta \mu=0$). For $N=3,4,5$ the Hamiltonian takes the form introduced in section~\ref{sec:Ham}, where the matrices for the layer degree of freedom $\gamma_i$ now are of dimension $N \times N$. The hopping term $t_\perp$ is kept constant for all $N$, see appendix~\ref{sec:AppHopp}. For the two types of pockets, the trilayer superconducting susceptibility is
\begin{align}
    \chi_{s, K}^{\text{3L}} & \approx  \left( \frac{1}{3} + \frac{2}{3} \frac{\left(\lambda_0^K \right) ^2}{(2 t_\perp^2 + \left(\lambda_0^K \right)  ^2)} \right) \chi_P \\ & \notag  + \frac{2}{3} \frac{2 t_\perp^2}{2 t_\perp^2 + \left(\lambda_0^K \right)  ^2}  \chi_{\text{sg}}(T) 
\end{align}
\begin{align}
    \chi_{s, \Gamma}^{\text{3L}}& \approx  \left( \frac{1}{3} + \frac{2}{3} \left( 1 -\frac{\sqrt{2} t_\perp }{\sqrt{2 t_\perp^2 + \left(\lambda_0^\Gamma \right) ^2} } \right) \right) \chi_P  \\ &\notag  + \frac{2}{3} \frac{\sqrt{2} t_\perp }{ \sqrt{2 t_\perp^2 + \left(\lambda_0^\Gamma \right) ^2} }  \chi_{\text{sg}}(T) 
\end{align}
and the susceptibility difference is
\begin{align}
   \delta \chi_{K}^{\text{3L}} & \approx  \frac{2}{3} \frac{2 t_\perp^2}{2 t_\perp^2 + \left(\lambda_0^K \right)  ^2} (\chi_P - \chi_{\text{sg}}(T) )\\ \delta \chi_\Gamma^{\text{3L}}& \approx  \frac{2}{3} \frac{\sqrt{2} t_\perp }{ \sqrt{2 t_\perp^2 + \left(\lambda_0^\Gamma \right)^2} } (\chi_P - \chi_{\text{sg}}(T) )
\end{align}
For $N=4$ the total inversion symmetry is preserved, and all bands are pair-wise degenerate:
\begin{align}
   \delta \chi_{K}^{\text{4L}} &\approx \frac{1}{2} \left( \frac{\left( \frac{3}{2} + \frac{\sqrt{5}}{2}\right) t_\perp^2}{\left( \frac{3}{2} + \frac{\sqrt{5}}{2}\right) t_\perp^2 + \left(\lambda_0^K \right)  ^2} \right. \\ &\notag \left. + \frac{\left( \frac{3}{2} - \frac{\sqrt{5}}{2}\right) t_\perp^2}{\left( \frac{3}{2} - \frac{\sqrt{5}}{2}\right) t_\perp^2 + \left(\lambda_0^K \right)  ^2} \right) (\chi_P - \chi_{\text{sg}}(T) )
\end{align}
\begin{align}
   \delta \chi_\Gamma^{\text{4L}}& \approx \frac{1}{2} \left( \frac{\sqrt{ \frac{3}{2} + \frac{\sqrt{5}}{2} } t_\perp}{ \sqrt{ \left(\frac{3}{2} + \frac{\sqrt{5}}{2}\right) t_\perp^2 + \left(\lambda_0^\Gamma \right) ^2}} \right. \\ &\notag \left.  + \frac{\sqrt{ \frac{3}{2} - \frac{\sqrt{5}}{2} } t_\perp}{ \sqrt{ \left(\frac{3}{2} - \frac{\sqrt{5}}{2}\right) t_\perp^2 + \left(\lambda_0^\Gamma \right) ^2}}\right) (\chi_P - \chi_{\text{sg}}(T) )
\end{align}
For 5 layers there is once again one set of spin-split bands:
\begin{align}
   \delta \chi_{K}^{\text{5L}} &\approx \frac{2}{5} \left( \frac{t_\perp^2}{ t_\perp^2 + \left(\lambda_0^K \right)  ^2} \right. \\ &\notag \left. + \frac{3 t_\perp^2}{3 t_\perp^2 + \left(\lambda_0^K \right)  ^2} \right) (\chi_P - \chi_{\text{sg}}(T) )
 \end{align}
 \begin{align}
\delta \chi_\Gamma ^{\text{5L}} &\approx \frac{2}{5} \left( \frac{ t_\perp}{ \sqrt{ t_\perp^2 + \left(\lambda_0^\Gamma \right) ^2}} \right. \\ &\notag \left. + \frac{\sqrt{3 } t_\perp}{\sqrt{ 3 t_\perp^2 + \left(\lambda_0^\Gamma \right) ^2}} \right) (\chi_P - \chi_{\text{sg}}(T) ) 
\end{align}
These expressions are valid when the multilayer has no bias potential, e.g., it is a free-standing sample. Like in section~\ref{sec:2Hbi}, this is used as a lower limit for the possible values of the critical field, and a theoretical upper limit is given by the scenario where each layer is considered as an isolated monolayer. The limits are compared to the experimentally measured multilayer $H_{c2}$ in Fig.~\ref{fig:allLayers}. As the critical temperature increases with the number of layers in NbSe$_2$, the window in which $H_{c2}$ can lay narrows. On the contrary, the window widens for TaS$_2$. Just like in the bilayer, the experimental data exceeds the lower limit by a fair amount.

In the bilayer case we modeled the effect from a substrate with an interlayer bias potential. However, when $N > 2$ there are multiple choices for which parameters to include in a model of how a substrate affects each layer. Such a modeling is beyond the scope of this work. However, the presence of such an effect should not be ruled out.

\section{Discussion}
By using a model which includes all pockets at the FS with parameters that closely reproduce the upper critical fields in the monolayer limit\cite{Engstrom2025}, we can establish some constraints on the possible symmetries of the superconducting order parameters in the few-layer compounds. 

From our calculations, an intralayer singlet order with a spin-triplet component, in an $s+f$-wave or $d+id$ and $p+ip$ symmetry, could be a likely order. This mixed-parity order would however result in a critical field of a similar magnitude to a purely spin-singlet order and would still require some interlayer bias potential to reproduce the experimental critical field. The same effect would be present if the order parameter has a nodal spin-singlet component that introduces an additional momentum-dependence\cite{Roy2026}. 

A purely spin-triplet order could only be compatible with the experimental data if the order changes sign between layers, as discussed in section~\ref{sec:phiPi}. Further, our proposed experiment in section~\ref{sec:Exp} would fully rule out a purely spin-triplet order if a scaling $H_{c2} \propto \sqrt{\delta \mu / t_\perp}$ were to be observed. We predict that experiments should show the predicted spin-singlet scaling independently of subleading effects, even if there is a large triplet component. 

Moreover, the rather large $\delta \mu $ that fits the data of the bilayers in Fig.~\ref{fig:biLayer}, for a spin-singlet, questions if some model parameter values should be modified to accurately reflect the physical system. Even so, a few cases can be ruled out as the sole explanation for the increased critical field. Firstly, the interlayer hopping at the $\Gamma$-pocket could potentially be smaller than expected. However, for bilayer NbSe$_2$ it would be required to be $t_\perp \approx 5.2$meV to reproduce the experimental data, a third of the expected value. Secondly, we have assumed that all pockets remain of equal size, see Eq.~\eqref{eq:Hc22H}. The critical field would increase if the proportion of states at the FS belonging to the $\Gamma$-pocket decreases. For this to be the effect which increases the critical field, it would require an unmotivated\cite{Hall2019} and significant decrease of the density of states for the pocket to $N_\Gamma (0)/ N_K (0) \approx 0.12$. 

Beyond the scope of this work, there is a possibility that the superconducting order develops a spin-triplet component once an in-plane field is applied, such as an $i d_y$-triplet component for a field along the $x$-direction\cite{Mockli2020}. This order is neither affected by the SOC nor significantly by the magnetic field, and if energetically favorable it would increase the critical field. 

For the 2H-stacking, monolayers might be significantly different from the few-layer systems, as the jump in $T_c$ could indicate a different paring mechanism\cite{DeLaBarrera2018,Das2023}. It should also be reiterated that for an increasing number of layers $N$ the $T_c$ increases for 2H-NbSe$_2$ while it decreases for 2H-TaS$_2$. For the model parameters utilized in this work, no qualitative difference is observed for the behavior of $H_{c2}/H_p$ between the two families of compounds.

When going to thicker samples, which become more 3D in nature, the orbital coupling of the magnetic field eliminates the differences between spin-symmetries in the response to the field. Further, this is also the case in twisted bilayer systems, see appendix~\ref{sec:AppTwist}. 

\section{Conclusions}
In this work we have shown that a 3-pocket model of few-layer 2H-TMDs provides a qualitatively consistent description of the experimental upper critical field data. As the susceptibility is strongly dependent on the SOC, the states closest to the symmetry-imposed SOC nodes are the, by far, most important for $H_{c2}$ in the $N$-layer compounds. Just like in the monolayer case, it is close to those points that different pairing symmetries will have the greatest difference in response.

The most likely intralayer spin-singlet order will, due to the presence of the $\Gamma$-pocket, have a unique scaling with SOC, which we predict to be observable by applying an increasing displacement field. However, we have shown that the scaling offers little information about subleading terms.

By comparing the predicted magnitude of the critical field for different dominating spin-symmetries in the possible mixed-parity orders, only a few options are incompatible with experimental data. Moreover, the accuracy in predicting the magnitude of the critical field suffers from the large number of model parameters that must be included. Our predicted experiment is one option for how this obstacle might be bypassed, by accessing the upper critical field's scaling with experimental tuning parameters.


\begin{acknowledgments}This work was supported by the French Agence Nationale de la Recherche (ANR), under grant number ANR-22-CE30-0037. We would also like to thank Tristan Cren and Ludovica Zullo for interesting discussions.
\end{acknowledgments}

\appendix

\section{Susceptibility in bilayers}\label{sec:AppSuscp}
At $h=0$, the eigenstates of the Hamiltonian in Eq.~\eqref{eq:HamSC} give us the normal state bands as $\xi_{\zeta l} = \xi_k + \zeta l \sqrt{t_\perp^2 + |\lambda_{\bk}|^2}$ with spin $\zeta$ and layer-index $l$. An intraband spin-singlet superconductivity is fully intraband with $E_{\zeta l} = \sqrt{\xi_{\zeta l}^2 + |\Delta|^2}$. At $h>0$ the normal state bands become $\xi_{h, \zeta l} = \xi_{\bk} + \zeta l \sqrt{(h + l t_\perp)^2 + |\lambda_{\bk}|^2}$. Here we can directly calculate:
\begin{align}
   \left. \frac{\partial \xi_{h, \zeta l}}{\partial h} \right|_{h=0} &=  \zeta l \frac{t_\perp }{ \sqrt{t_\perp^2 + |\lambda_{\bk}|^2}}, \\ \left. \frac{\partial^2 \xi_{h, \zeta l}}{\partial h^2} \right|_{h=0} &= \zeta \frac{|\lambda_{\bk}|^2}{(t_\perp^2 + |\lambda_{\bk}|^2)^{3/2}}.
\end{align}
We have the intraband and the interband contributions to the normal state susceptibility\cite{Sigrist2009} $\chi_N = \chi_N^{\text{intra}}  + \chi_N^{\text{inter}}$, with
\begin{align}
    \chi_N^{\text{intra}} = &\sum_{\zeta l, \bm{k}} \left( \left. \frac{\partial \xi_{h, \zeta l}}{\partial h} \right|_{h=0} \right)^2 \frac{\partial f(\xi_{\zeta l})}{\partial \xi_{\zeta l}}~,\\
    \chi_N^{\text{inter}} = &\sum_{\zeta l, \bm{k}} \left. \frac{\partial^2 \xi_{h, \zeta l}}{\partial h^2} \right|_{h=0}  f(\xi_{\zeta l}),
\end{align}
where $f$ is the Fermi-Dirac distribution at temperature $T$. The limits of the integration range is set to $k_\text{min}^2 = \frac{\mu - \epsilon_\text{min}}{\alpha} $ and $k_\text{max}^2 = \frac{\mu - \epsilon_\text{max}}{\alpha} $, where we assume $\epsilon_\text{min}=- \epsilon_\text{max}=-\epsilon$ and $\epsilon > \lambda_0$. As previously established in inversion-symmetry-broken superconductors\cite{Skurativska2021,Maruyama2012} both intra- and inter-band components are non-zero and in this case evaluated as:
\begin{align}\label{eq:chiNintraTh}
\chi_{N}^{\text{intra}}(T)  &=\\&  \sum_{\zeta l}  \frac{N_{\zeta l} (0)}{2 \pi}  \int d \theta \int_{k_\text{min}}^{k_\text{max}} dk k \frac{t_\perp^2 }{ t_\perp^2 + |\lambda (\theta)|^2} \frac{\partial f(\xi_{\zeta l})}{\partial \xi_{\zeta l}} \notag \\ \notag
 &=   \sum_{\zeta l}  \frac{N_{\zeta l} (0)}{2 \pi}  \int d \theta \frac{t_\perp^2 }{ t_\perp^2 + |\lambda (\theta)|^2} \\ & \notag \times \int_{\epsilon_\text{min}}^{\epsilon_\text{max}}   d \xi_k \frac{\partial f(\xi_{\zeta l})}{\partial \xi_{\zeta l}} \\  &=   \frac{N(0)}{2 \pi}  \int d \theta \frac{t_\perp^2 }{ t_\perp^2 + |\lambda (\theta)|^2} \notag
\end{align}
\begin{align}\label{eq:chiNinterTh}
\chi_{N}^{\text{inter}}(T)  &=\\&  \notag \sum_{\zeta l}  \frac{N_{\zeta l} (0)}{2 \pi}  \int d \theta \int_{k_\text{min}}^{k_\text{max}} dk k  \frac{\zeta l |\lambda (\theta)|^2 f( \xi_{\zeta l}) }{(t_\perp^2 + |\lambda (\theta)|^2)^{3/2}}  \\ \notag
 &= \frac{N(0)}{2 \pi}  \int d \theta \frac{|\lambda (\theta)|^2}{t_\perp^2 + |\lambda (\theta)|^2} \frac{ T }{|\eta (\theta)|}\\ \notag &\times  \ln \left(\frac{ f ( \epsilon_\text{max} + |\eta (\theta)|) f ( \epsilon_\text{min} - |\eta (\theta)|)}{f ( \epsilon_\text{max} - |\eta (\theta)|) f ( \epsilon_\text{min} + |\eta (\theta)|)} \right) \\ \notag &=  \frac{N(0)}{2 \pi}  \int d \theta \frac{|\lambda (\theta)|^2}{t_\perp^2 + |\lambda (\theta)|^2} \\ &=  \frac{\chi_P}{4 \pi}  \int d \theta \frac{|\lambda (\theta)|^2}{t_\perp^2 + |\lambda (\theta)|^2},\notag
\end{align}
where $|\eta (\theta)| = \sqrt{t_\perp^2 + |\lambda (\theta)|^2}$. 
If we now evaluate Eq.s~\eqref{eq:chiNintraTh} \&~\eqref{eq:chiNinterTh} for each type of pocket, we get
\begin{align}
\chi_{N}^{K, \text{intra}}(T)  = \frac{t_\perp^2 }{ t_\perp^2 + \left( \lambda_0^K \right)^2} \chi_P ,\\ 
\chi_{N}^{K, \text{inter}}(T)  = \frac{\left( \lambda_0^K \right)^2 }{ t_\perp^2 + \left( \lambda_0^K \right)^2} \chi_P,
\end{align}
for $K$-pockets with $\lambda(\theta) = \lambda_0^K$ and so for a $\Gamma$-pocket with $\lambda(\theta) = \lambda_0^\Gamma \cos 3 \theta$:
\begin{align}
\chi_{N}^{\Gamma, \text{intra}}(T)  &= \frac{t_\perp }{\sqrt{ t_\perp^2 + \left( \lambda_0^\Gamma \right)^2} } \chi_P, \\ \chi_{N}^{\Gamma, \text{inter}}(T)  &= \left( 1- \frac{ t_\perp }{\sqrt{ t_\perp^2 + \left( \lambda_0^\Gamma \right)^2} } \right) \chi_P.
\end{align}
Note that in both cases we recover the expected $\chi_N = \chi_P$.

When considering an intraband spin-singlet order in the 2H-stacked bilayer that conserves inversion symmetry the superconducting order parameter can be considered a small parameter, in which we expand the bands:
\begin{align}
    E_{h, \zeta l} &\approx  | \xi_{h, \zeta l} | +\\ \notag  &\frac{\Delta ^2 \left( h (h+ l t_\perp) + \zeta l \xi_{\bk}  \sqrt{(h+ l t_\perp)^2+|\lambda_{\bk}| ^2} \right)}{X_{h,\zeta l}},
\end{align}
with
\begin{align}
    X_{h,\zeta l} &= 2 h^2 \xi_{\bk}  + 2 \zeta l  \xi_{\bk}  ^2 \sqrt{(h+ l t_\perp)^2+|\lambda_{\bk}| ^2} \\ \notag &+2 h t_\perp l \left( 2 \xi_{\bk} + \zeta l \sqrt{(h+ l t_\perp)^2+|\lambda_{\bk}| ^2}  
   \right) \\ &+ 2 \xi_{\bk}  \left(|\lambda_{\bk}| ^2+t_\perp^2\right),\notag
\end{align}
and therefore
\begin{align}
   \left. \frac{\partial E_{h, \zeta l}}{\partial h} \right|_{h=0} &\approx \left. \frac{\partial \xi_{h, \zeta l}}{\partial h} \right|_{h=0}, \\ \left. \frac{\partial^2 E_{h, \zeta l}}{\partial h^2} \right|_{h=0} &\approx \left| \left.\frac{\partial^2 \xi_{h, \zeta l}}{\partial h^2} \right|_{h=0} \right| + \mathcal{O} \left( \Delta^2 \right).
\end{align}
The susceptibility in the superconducting state is calculated as\cite{Samokhin2021}:
\begin{equation}
    \chi_S^{\text{intra}} = \sum_{\zeta l, \bm{k}} \left( \left. \frac{\partial E_{h, \zeta l}}{\partial h} \right|_{h=0} \right)^2 \frac{\partial f(E_{\zeta l})}{\partial E_{\zeta l}},
\end{equation}
and $\chi_{S}^{\text{inter}}= \chi_{S,T}^{\text{inter}} + \chi_{S,0}^{\text{inter}}$ with
\begin{align}\label{eq:chiSinterT}
    \chi_{S,T}^{\text{inter}} =& \sum_{\zeta l, \bm{k}} \left. \frac{\partial^2 E_{h, \zeta l}}{\partial h^2} \right|_{h=0}  f(E_{\zeta l}) 
\end{align}
\begin{align}\label{eq:chiSinter0}
    \chi_{S,0}^{\text{inter}} =& \sum_{\zeta l, \bm{k}}  \frac{1}{2}\left( \left. \frac{\partial^2 \xi_{h, \zeta l}}{\partial h^2} \right|_{h=0} - \left. \frac{\partial^2 E_{h, \zeta l}}{\partial h^2} \right|_{h=0} \right).
\end{align}
We evaluate this expression for the bilayer as
\begin{align}
\chi_{s,T}^{\text{intra}}(T)& =  \\ \notag &\sum_{\zeta l} \frac{N_{\zeta l}(0)}{2 \pi} \int d \theta d \xi_k \frac{ \zeta l \xi_{\zeta l}  f( E_{\zeta l}) }{|\eta (\theta)| \sqrt{\xi_{\zeta l}^2} }  + \mathcal{O} \left( \Delta^2 \right) \\ \notag
&= \frac{N(0)}{2 \pi} \int d \theta \frac{ T }{|\eta (\theta)|} \ln \left(\frac{ f ( E_{+,\text{max}}) f ( E_{-,\text{min}})}{f ( E_{-,\text{max}}) f ( E_{+,\text{min}})} \right)   \\ \notag& + \chi_{S,0} +  \mathcal{O} \left( \Delta^2 \right)\\ \notag& =\chi_{S,0} +  \mathcal{O} \left( \Delta^2 \right),
\end{align}
and
\begin{align}
\chi_{S,0}^{\text{inter}}(T)  \approx\frac{1}{2}  \frac{1}{2 \pi}   \int d \theta \frac{|\lambda (\theta)|^2 }{ t_\perp^2 + |\lambda (\theta)|^2} \chi_P + \mathcal{O} \left( \Delta^2 \right).
\end{align}
Therefore,
\begin{align}
\chi_{S}^{\text{intra}}(T) &= \frac{1}{2 \pi}  \int d \theta \frac{t_\perp^2 }{ t_\perp^2 + |\lambda (\theta)|^2} \\ 
& \notag \times\sum_{\zeta l} N_{\zeta l}(0)  \int d \xi_{\bm{k}} \frac{\partial f( E_{\zeta l})}{\partial E_{\zeta l}}  \notag \\ \notag
&\approx \frac{\chi_P}{2 \pi}  \int d \theta \frac{t_\perp^2 }{ t_\perp^2 + |\lambda (\theta)|^2}  \int d \xi_{\bm{k}} \frac{\partial f( E_{k})}{\partial E_{k}},
\end{align}
and by introducing the Yoshida function $Y(T)$, this expression can be written as
\begin{align}
\chi_{S}^{\text{intra}}(T)  \approx   \frac{1}{2 \pi}  \int d \theta \frac{t_\perp^2 }{ t_\perp^2 + |\lambda (\theta)|^2} Y(T)  \chi_P 
\end{align}
\begin{align}
\chi_{S}^{\text{inter}}(T)  \approx \frac{1}{2 \pi}   \int d \theta \frac{|\lambda (\theta)|^2 }{ t_\perp^2 + |\lambda (\theta)|^2} \chi_P + \mathcal{O} \left( \Delta^2 \right).
\end{align}
The susceptibility difference is thus 
\begin{align}
\delta \chi^{K}(T)  &=\chi_{N}^{K}(T) - \chi_{S}^{K}(T)   \\ \notag &= \frac{t_\perp^2 }{ t_\perp^2 + \left( \lambda_0^K \right)^2} \left( 1 - Y(T) \right) \chi_P + \mathcal{O} \left( \Delta^2 \right),
\end{align}
\begin{align}
\delta \chi^{\Gamma}(T)  &=\chi_{N}^{\Gamma}(T) - \chi_{S}^{\Gamma}(T)   \\ \notag &=  \frac{t_\perp }{\sqrt{ t_\perp^2 + \left( \lambda_0^\Gamma \right)^2} }  \left( 1 - Y(T) \right) \chi_P + \mathcal{O} \left( \Delta^2 \right).
\end{align}
For the compounds considered in this work we are considering ratios such as $\Delta / \lambda_0^\Gamma \approx 0.04$ while $t_\perp / \lambda_0^\Gamma \approx 0.9$, meaning that the $\mathcal{O} \left( \Delta^2 \right)$-terms can be neglected.

\section{Validity of the 3-pocket model}\label{sec:dissVal}

There is a physical motivation for including the $\Gamma$-pockets to explain the critical field found in experiment, as the interlayer coupling is stronger for pockets of this type. Furthermore, the 3-pocket model $H_{c2}^3(T)$ provides a more consistent prediction of $H_{c2}/H_p$ in the bilayers than previous models that have only considered the $K$-pockets via $H_{c2}^K(T)$. To compare the two models we define:
\begin{align}
    H_{c2}^K(T) = \sqrt{\frac{\Omega^K_0 (T)}{\delta \chi^K (T)}}, \\
     H_{c2}^3(T) = \sqrt{\frac{\Omega^\Gamma_0 (T) + \Omega^K_0 (T)}{\delta \chi^\Gamma (T)+  \delta \chi^K (T)}}.
\end{align}
 Considering the bilayer with preserved inversion symmetry, for either the three pockets or for only the $K$-pockets, Eq.~\eqref{eq:2LKG}, results in
\begin{align}\label{eq:Hc22H}
   \frac{H_{c2}^K(0)}{H_p} \approx  \frac{ \sqrt{t_\perp^2 + \left(\lambda_0^K\right) ^2}}{ t_\perp}\\ \frac{H_{c2}^3(0)}{H_p} \approx \sqrt{ \frac{N_\Gamma (0) +N_K (0)}{N_\Gamma (0)} \frac{ \sqrt{t_\perp^2 + \left(\lambda_0^\Gamma\right) ^2}}{ t_\perp}}
\end{align}
with the density of states at the FS $N_K (0)=2 N_\Gamma (0)$. The results in Fig.~\ref{fig:biLayer} can be compared to Ref.~\cite{DeLaBarrera2018} who, by using only the $K$-pockets for bilayer 2H-NbSe$_2$, find $\frac{H_{c2}^K(0)}{H_p} \approx 3.5 $. This value is in fact a closer match to the experimental value than what we find for the 3-pocket model $\frac{H_{c2}^3(0)}{H_p} \approx 2$ (the lower limit in Fig.~\ref{fig:biLayer}). On the other hand, using only $K$-pockets predicts $\frac{H_{c2}^K(0)}{H_p} \approx 17.5 $ for bilayer 2H-TaS$_2$, almost three times the experimental value, in contrast to the 3-pocket model that fits much better with $\frac{H_{c2}^3(0)}{H_p} \approx 4$.

Even though the 3-pocket model provides more consistent results, it underestimates the values of the critical field (further shown to be the case for $N$-layers in section~\ref{sec:N2multi}). If a slight overestimation compared to experimental data had been found, it might not have indicated qualitative problems with the model, as factors that decrease the critical field are expected to be always present. Among these factors are disorder or a small Rashba SOC\cite{Mockli2020,Ilic2017}. The underestimation that we find, on the contrary, indicates that some additional effect must be included in the model. We have suggested an origin in section~\ref{sec:effMono}, attributing the increase of the critical field to an inversion-symmetry-breaking in experiments.

\begin{figure}
    \centering
    \includegraphics[width=\linewidth]{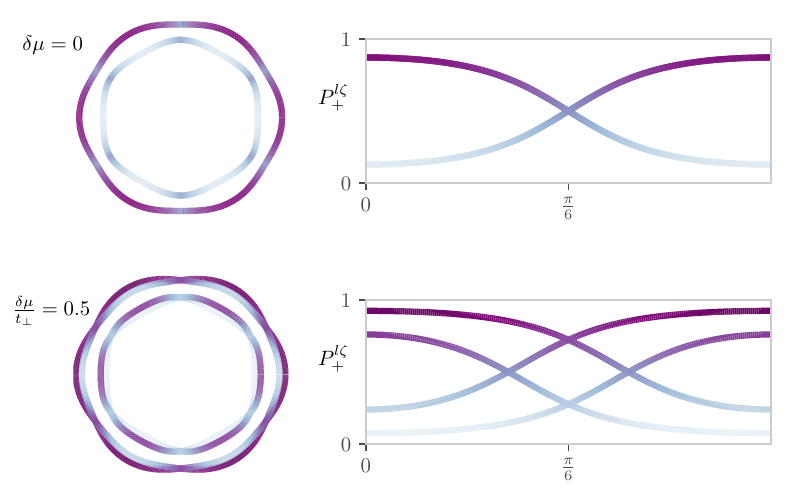}
    \caption{\label{fig:FSGamma}Normal state FS for the $\Gamma$-pocket at $h =0$, with layer-polarization $P_{+}^{\zeta l}$. At $\delta \mu=0$ the bands are doubly degenerate and the spin-up copies of the bands are shown, where the spin-down bands have the opposite layer-polarization.}
\end{figure}
\section{Eigenstates with interlayer bias potential}\label{sec:AppEigs}
When an interlayer bias potential is present the normal state bands for the $\Gamma$-pocket, the spin- and layer-polarization of the eigenstates will have a different character depending on the distance to the SOC nodes. At $h=0$ the bands in Eq.~\eqref{eq:eigDmu} have the four eigenstates 
\begin{align}
b_{\zeta l} &= \left( \left( (l \delta \mu +  \zeta \lambda_{\bk})  + \zeta l \sqrt{t_\perp^2 +  (l \delta \mu + \zeta\lambda_{\bk})  }\right) c_{ \zeta +} \right. \\ \notag
 &\left. + t_\perp c_{ \zeta -} \right) / \mathcal{N}_{\zeta l},
\end{align}
for layer index $l$ and spin $\zeta$. In Fig.~\ref{fig:FSGamma} the projection onto the top layer ($+$):
\begin{align}
    P_{+}^{\zeta l} &= \langle c^\dagger _{ \zeta +} b^\dagger_{\zeta l} b_{\zeta l}   c _{\zeta +}  \rangle \\ \notag &= \frac{l \delta \mu + \zeta \lambda_{\bk} + \zeta l \sqrt{(l \delta \mu + \zeta \lambda_{\bk})^2 + t_\perp^2}}{ 2 \sqrt{(l \delta \mu + \zeta \lambda_{\bk})^2 + t_\perp^2}},
\end{align}
is shown at $\delta \mu=0$ and $\delta \mu \neq 0$. At $\delta \mu=0$ the mixing between layers increases as the ratio $t_\perp / \lambda_{\bk}$ decreases. Even at the furthest point from the nodes, the mixing remains significant. At $\delta \mu \neq 0$ the states exhibit a distinct character close by the nodes, where the SOC can be considered small (see Eq.~\eqref{eq:eigsDmuLa}), and far from the nodes, where the limit of small $\delta \mu$ can be considered:
\begin{align}
    \xi_{\zeta l} \approx \xi_{\bm{k}} + \zeta l\sqrt{ |\lambda_{\bm{k}}|^2 +t_\perp^2} + \zeta   \frac{\delta \mu |\lambda_{\bm{k}}|}{\sqrt{ |\lambda_{\bm{k}}|^2 +t_\perp^2}}.
\end{align}
Close by the nodes the bands that intersect have the same layer polarization. Far from the nodes the layer polarization becomes stronger than at $\delta \mu=0$ and instead resembles having isolated monolayers. The effective SOC defined from Eq.~\eqref{eq:eigsDmuLa} is thus valid for states close to the SOC nodes. As these are the states with the largest contribution to the susceptibility, it explains the close overlap of the effective model with the full numerical evaluation of the susceptibility in section~\ref{sec:dmu}. It should however be noted that a monolayer model with an effective SOC is not expected to capture the physics of the bilayer when evaluating other quantities.

\section{Interpolated effective model}\label{sec:AppInter}
When the interlayer bias potential $\delta \mu$ is non-zero at the $\Gamma$-pocket the FS has degenerate states at the SOC node. We will utilize that for a singlet order the states close to a node contribute, by far, the most to the susceptibility difference between the normal and superconducting state. 

In an attempt to bridge the two limits $\delta \mu =0$ and $\delta \mu \rightarrow \infty$, we assume that close to a SOC node $\lambda(\theta_0)=0$ there are two contributions, one from the intraband-interband difference of the normal state susceptibility and the other from the effective monolayer model with an effective SOC. Both terms are assumed to exist for any value of $\delta \mu$. Expanding in small $|\lambda_{\bk}(\theta \approx \theta_0) | = v_\lambda (\theta - \theta_0)$:
\begin{align}\label{eq:interPolExp}
    \delta \chi_{\delta \mu} (\theta \approx \theta_0) \approx \delta \chi_{\delta\mu=0} (\theta \approx \theta_0) + \delta \chi_{\frac{\delta\mu}{t_\perp} \approx 1} (\theta \approx \theta_0), \\ \notag
    = 1 - \left(  \frac{\pi^2}{3} \frac{1}{t_\perp^2} +   \frac{\delta\mu^2}{\delta \mu^2 + t_\perp^2}\frac{1}{\Delta_0^2} \right) v_\lambda^2 (\theta - \theta_0)^2.
\end{align}
As in both limits $\delta \chi (\theta)$ has an approximately Lorentzian form, and we assume that $\delta \chi_{\delta \mu}  (\theta)$ can be described as a single Lorentzian $ \propto L(\theta, \alpha_0)$\cite{Engstrom2025}. From the expansion close to the node Eq.~\eqref{eq:interPolExp}, the proportionality constants give us
 \begin{align}
    \frac{\delta \chi_{\delta \mu} (\theta )}{\chi_P} &\approx \left(  \left( \frac{ \lambda_0^\Gamma}{t_\perp} \right)^2  \right. \\ \notag &\left.+ \left( \frac{\sqrt{3}}{\pi}\right)^2 \frac{\delta\mu^2}{\delta \mu^2 + t_\perp^2} \left( \frac{\lambda_0^\Gamma}{\Delta_0}\right)^2 \right)^{-\frac{1}{2}} L(\theta, \alpha_0).
\end{align}
After integration over $\theta$ we can express the interpolated effective model as:
 \begin{align}\label{eq:interPolHc2}
    \frac{H_{c2}(\delta \mu)}{H_P} \approx \left( \left( \frac{ \lambda_0^\Gamma}{t_\perp} \right)^2 +  \left( \frac{\sqrt{3}}{\pi}\right)^2 \frac{\delta\mu^2}{\delta \mu^2 + t_\perp^2} \left( \frac{ \lambda_0^\Gamma}{\Delta_0} \right)^2 \right)^{\frac{1}{4}}.
\end{align}
The critical field Eq.~\eqref{eq:interPolHc2} is shown in Fig.~\ref{fig:dmu_var}, where it follows the curve of the numerical integration of the full susceptibility. The small difference between the two methods at intermediate $\delta \mu$, might come from the Lorentzian approximation or additional scaling ratios that for the given values remain close to one.

At intermediate to large $\delta \mu$ the effective model scales, as expected, as $\sqrt{\frac{\delta \mu}{t_\perp}}$ while at very small $\delta \mu \rightarrow 0$ we find the scaling:
\begin{align}
     \frac{H_{c2}(\delta \mu\approx0)}{H_P} \approx \frac{H_{c2}(0)}{H_P} \left( 1 + \frac{3}{4 \pi^2} \left( \frac{\lambda_0^\Gamma}{\Delta_0}\right)^2 \left(  \frac{\delta \mu}{ \lambda_0^\Gamma} \right)^2 \right).
\end{align}

\section{Mixed-parity susceptibility}\label{sec:AppMixed}
To understand the contributions to a mixed-parity susceptibility in the monolayer, we consider the limit of a small $\Delta_{t,0}/\Delta_{s,0} \ll 1$. Eq.~\eqref{eq:mixMonoDer} becomes 
\begin{align}
  D^{\text{mix}}_{\zeta}= \frac{1}{\mu_{B}^2} \left. \frac{\partial^2 E_{h, \zeta}^{\text{mix}}}{\partial h^2} \right|_{h=0} \approx D^{\text{sing}}_{\zeta} + D^{(1)}_{\zeta}  + D^{(2)}_{\zeta},
\end{align}
where $D^{\text{sing}}_{\zeta} = \frac{\zeta l}{|\lambda_{\bm{k}}|} \left( \frac{\xi_{l,\zeta}}{E_{l,\zeta}^{\text{sing}}} + \frac{\Delta}{\xi_{\bm{k}} E_{l,\zeta}^{\text{sing}}}\right)$ is the term for a spin-singlet in a monolayer, the term $D^{(1)}_{\zeta}$ vanishes under integration for the two spin species, and 
\begin{align}
    D^{(2)}_{\zeta} =&  \frac{|\Delta_t|^2}{\left( E_{l,\zeta}^{\text{sing}}\right)^3 }  \\ & \notag \times
    \left(\frac{|\Delta_s|^2 \left( \xi_{\bk}^2 + |\Delta_s|^2 \right)}{\left( \xi_{\bk} \lambda_{\bk}\right)^2 } \left(1 + \zeta \frac{\left(E_{l,\zeta}^{\text{sing}} \right)^2 }{ |\xi_{\bk} | |\lambda_{\bk}|} \right) \right. \\ &\notag
\left. + \frac{ \xi_{\bk}^2 + |\Delta_s|^2 + \zeta |\xi_{\bk} | |\lambda_{\bk}|  }{2  |\xi_{\bk} | |\lambda_{\bk}|  }\left( \zeta  + \frac{|\Delta_s|^2}{\left(E_{l,\zeta}^{\text{sing}} \right)^2 } \right) \right).
\end{align}
Considering this second order term in $\Delta_t$ and expanding close to a SOC node at $\theta = \theta_0$, we use that $\lambda^\Gamma (\theta) \approx v_\lambda (\theta- \theta_0)$ and $\Delta_t \approx v_{\Delta_t} (\theta- \theta_0)$ if the pairing symmetry of the triplet is chosen to have coinciding nodes with the SOC. The expansion results in:
\begin{align}
   \sum_\zeta D^{(2)}_{\zeta} \approx 18 |\Delta_t|^2 \frac{\xi_{\bk}^2}{\left( \xi_{\bk}^2 + |\Delta_s|^2\right)^{\frac{5}{2}}} (\theta- \theta_0) ^2,
\end{align}
and after integrating the parabolic dispersion $\xi_{\bk}$ over the interval $\left[ -\epsilon, \epsilon\right]$ the contribution to the susceptibility for the superconducting state is
\begin{align}
    \chi_S^{(2)} (\theta \approx \theta_0) \approx  \frac{N_\Gamma(0)}{2 \pi} 6 \frac{|\Delta_t|^2}{|\Delta_s|^2} \frac{\epsilon^3 }{ \left( \epsilon^2 + |\Delta_s|^2\right)^{\frac{3}{2}}} (\theta- \theta_0) ^2.
\end{align}
If $|\Delta_s|/ \epsilon \rightarrow 0$ this is simply $\chi_S^{(2)} (\theta) \approx  \frac{N_\Gamma(0)}{2 \pi}  6 \frac{|\Delta_t|^2}{|\Delta_s|^2} (\theta- \theta_0) ^2$. 

We can compare this to the term acquired from the singlet pairing $D^{\text{sing}}_{\zeta}$\cite{Engstrom2025}: $\chi_S^{{\text{sing}}} (\theta \approx \theta_0) \approx \frac{N_\Gamma(0)}{2 \pi}  6 \frac{|\lambda_0^\Gamma|^2}{|\Delta_s|^2} (\theta- \theta_0) ^2$. As we consider the limits $\lambda_0^\Gamma \gg |\Delta_s|, |\Delta_t|$ the contribution to the susceptibility in the superconducting state will be much larger from the singlet order than from the term containing the triplet order parameter $\chi_S^{\text{sing}} (\theta \approx \theta_0) \gg \chi_S^{(2)} (\theta \approx \theta_0)$. Assuming that we can treat $\delta \chi^\Gamma (\theta)$ as a Lorentzian, supported by Fig.~\ref{fig:mixParity}, the upper critical field has the form
\begin{equation}
    \frac{H_{c2}^\Gamma}{H_p} \propto \sqrt{\frac{\lambda_0^\Gamma}{\Delta_{s,0}}} \left( 1 + \left( \frac{\Delta_{t,0}}{\lambda_0^\Gamma} \right)^2 \right)^\frac{1}{4} .
\end{equation}
As expected, the mixed-parity critical field is primarily determined by the spin-singlet component.

\section{Multi-layer hopping term}\label{sec:AppHopp}
Concerning the hopping parameters used in Fig.~\ref{fig:allLayers}, we have chosen to keep $t_\perp$ constant as the number of layers increases. As shown in Ref.~\cite{DeLaBarrera2018} the total splitting between bands, as calculated by DFT, increases with number of layers until it reaches a bulk value. We can define an effective hopping parameter $\tilde{t}_{\perp, N}$, which determines such a splitting at $\lambda_{\bk} =0$. This quantity increases with the number of layers, as for bi- and tri-layers we observe that
\begin{equation}
    \tilde{t}_{\perp, 2} = t_{\perp}, \qquad \tilde{t}_{\perp, 3} = \sqrt{2} t_{\perp}.
\end{equation}

\section{Twisted bilayer NbSe$_2$}\label{sec:AppTwist}
In Ref.~\cite{Zhong2025} a bilayer of two twisted monolayers of NbSe$_2$, with a small twisting angle around $1^\circ$, have an in-plane upper critical field with a roughly linear temperature dependence. Given the critical temperature the Pauli limit is $H_p \approx 13$T and using a linear extrapolation of the data results in $H_{c2}(0)/H_p \approx 1.38$. The critical temperatures are taken for the resistivity data half-way through the transition. The linear temperature dependence is a clear signature that the orbital coupling of the magnetic field is dominant\cite{Tinkham1962,Klemm1975}.

We do not know what topology the FS has in the twisted system. However, the structure differs from the 2H-stacking as both layers have the same orientation, in which they break inversion symmetry. Assuming that some nodes of the SOC remain at the FS, we can consider the value of the susceptibility from a $\Gamma$-like pocket with $\frac{\lambda_0^\Gamma}{\Delta_0} \approx 15.7$. If the full susceptibility must result in the experimental value of the critical field 
\begin{equation}
    \frac{H_{c2}(T)}{H_p} = \sqrt{\frac{1}{\frac{\delta \chi^{\text{orb}}(T)}{\chi_P} + \frac{\delta \chi^{\text{Z}}(T)}{\chi_P}}}.
\end{equation}
it is required that $ \delta \chi^{\text{orb}}(T) \approx 10 \delta \chi^{Z} (T)$. Even though the twisted system is a bilayer, the greatly increased size of the unit cell makes the orbital coupling the dominant term.

\bibliography{multiRefs}

\end{document}